\documentclass[pre,aps,floatfix,letterpaper,reprint]{revtex4-1}

\usepackage{graphicx}
\usepackage{amsmath}
\usepackage{float}
\usepackage{placeins}
\usepackage{bm}
\usepackage{ulem}
\usepackage{xcolor}

\definecolor{green}{RGB}{0,150,0}

\begin{document}
%proposed title : 
% Interface-mixing driven Rayleigh-Taylor instability at the origin of granular axial segregation bands in rotating tumbler
%or Mixture driven RTI ...
\title{
Origin of granular axial segregation bands in a rotating tumbler: an interface-mixing driven Rayleigh-Taylor instability}

\author{Umberto D'Ortona}
\email{umberto.dortona@cnrs.fr}
\affiliation{Aix-Marseille Universit\'e, CNRS, Centrale Marseille, M2P2, Marseille, France}
\author{Richard M. Lueptow}
\affiliation{Department of Mechanical Engineering, 
         Northwestern University, Evanston, Illinois 60208, USA}
\affiliation{Department of Chemical and Biological Engineering, Northwestern University, Evanston, Illinois 60208, USA}
\author{Nathalie Thomas} 
\affiliation{Aix-Marseille Universit\'e, CNRS, IUSTI, Marseille, France}

\date{\today}
\begin{abstract}

The origin of large and small particle axial bands in long rotating tumblers is a long-standing question. Using DEM simulations, we show that this axial segregation is due to a Rayleigh-Taylor instability %, which can occur in a bidisperse granular medium because it has the property that packing, and, consequently, density, is higher for mixed particles than monodisperse particles. 
which is characterized by the fact that the density of a granular medium increases with mixing and decreases with segregation.
For initially mixed particles, segregation and collisional diffusion in the flowing layer balance and lead to a three-layer system, with a layer of large particles over a layer of small particles, and, interposed between these layers, a layer of more densely %bulk density 
packed mixed particles. The higher density  mixed particle layer over the lower density small  particle layer induces a Rayleigh-Taylor instability, evident as waviness in the interface between the layers.  The waviness destabilizes into ascending plumes of small particles and descending plumes of mixed particles with large particles enriched near the surface, which become evident as small and large particle bands visible at the free surface. Rolls driven by segregation at the tilted interface between plumes maintain the pattern of frozen plumes.

%\textbf{UMBERTO'S ORIGINAL VERSION:} The origin of large and small particle axial bands in long rotating tumblers is a long-standing question. Using DEM simulations, we show that axial segregation is due to a granular Rayleigh-Taylor instability in which {\blue the mixing between large and small particles forms the density excess, that can disappear by size segregation.  In an initially homogeneous medium, segregation and collisional diffusion spontaneously induce the formation of a three layer system, with a  layer of large particles over a layer of small particles, and in between, a layer of higher bulk density of mixed particles.} The higher bulk density layer over lower bulk density pure small  particle layer induces a Rayleigh-Taylor instability, evident as waviness in the interface between the layers.  The waviness destabilizes into ascending plumes of small particles and descending plumes of mixed particles with large particles enriched near the surface, which become evident as small and large particle bands visible at the free surface. Rolls driven by segregation at the tilted interface between plumes {\blue balance the plumes progression and} maintain the pattern.
\end{abstract}
\maketitle

%\section{Introduction}

\emph{Introduction}---{Rayleigh-Taylor (RT) instabilities arise when a denser fluid is placed atop a lighter one in a gravitational field, and the horizontal interface between the layers becomes wavy~\cite{Chandrasekhar61,Charru}. The amplitude of the waviness increases, eventually resulting in downward high density plumes and upward low density plumes. In its simplest form, the density difference originates from the use of distinct fluids, either miscible or immiscible. %\sout{Similarly, temperature variations can lead to a density difference, giving rise to the Rayleigh-Bénard instability \cite{Benard00}.}
However, RT instabilities
are also observed in more complex situations including chemical reactions at an interface forming a third layer with a density difference \cite{CareyMorris96,BockmannMuller00, MartinRako02},  the dissolution of a gas in a liquid \cite{Haut}, and the accumulation of particles \cite{AbkarianProtiere13,ProtiereJosserand17} or bubbles \cite{ThomasTait93} at the interface of two liquids. In these last examples, a third layer, denser or lighter, can appear in between the two initial layers creating a 3-layer RT instability, which has been addressed theoretically \cite{Matthews88,ListerKerr89,BatchelorNitsche91}. % and the instability growth is associated to a thickening of the third phase layer due to a horizontal displacement of its fluid. 
A special situation of interest for this study occurs in some of these systems (autocatalytic reaction front \cite{BockmannMuller00}, granular rafts \cite{ProtiereJosserand17}, CO$_2$ chemisorption \cite{Haut}, or simply a single thin layer of liquid bounded by an upper wall \cite{YiantsiosHiggins89}), where the RT instability initiates the interfacial waviness and then ascending or descending plumes start to form. But for different reasons, the plumes do not evolve further, so that a frozen pattern of protrusions results instead of the ongoing plumes that usually cause the eventual total reversal of the dense and light layers following the RT instability.} 

Rayleigh-Taylor instabilities can also occur in granular flows. 
Unlike the fluid instability,
the granular layers must be flowing for the instability to occur (static granular layers would remain unchanged without the input of energy via the flow). For example, for flow down a rough inclined chute with a layer of large dense particles above a layer of small light particles, the interface between the layers destabilizes to form ascending plumes of the small light particles and descending plumes of large dense particles in a spanwise-depthwise cross-section of the flow~\cite{DOrtonaThomas20}. These plumes appear at the surface as parallel stripes of the two particle species extending in the streamwise direction, and they can be sustained by counter-rotating Rayleigh-B\'enard-like convection rolls between the plumes { where the convection rolls are driven by granular size segregation.} 
This granular RT instability can even occur with initially mixed particles flowing down an incline: large dense particles initially segregate upward so that they accumulate in a dense particle layer that destabilizes to the plume pattern. { Because the unstable state forms spontaneously as the granular material is flowing, it has been termed self-induced RT instability~\cite{DOrtonaThomas20}. }
The objective
of this letter is to explain how 
a granular Rayleigh-Taylor instability causes the axial segregation of particles having two sizes, even though the particles have the same density, which may seem paradoxical.

\begin{figure}[hptb] %fig 1 % equivalent to 263words
\includegraphics[width=\linewidth]{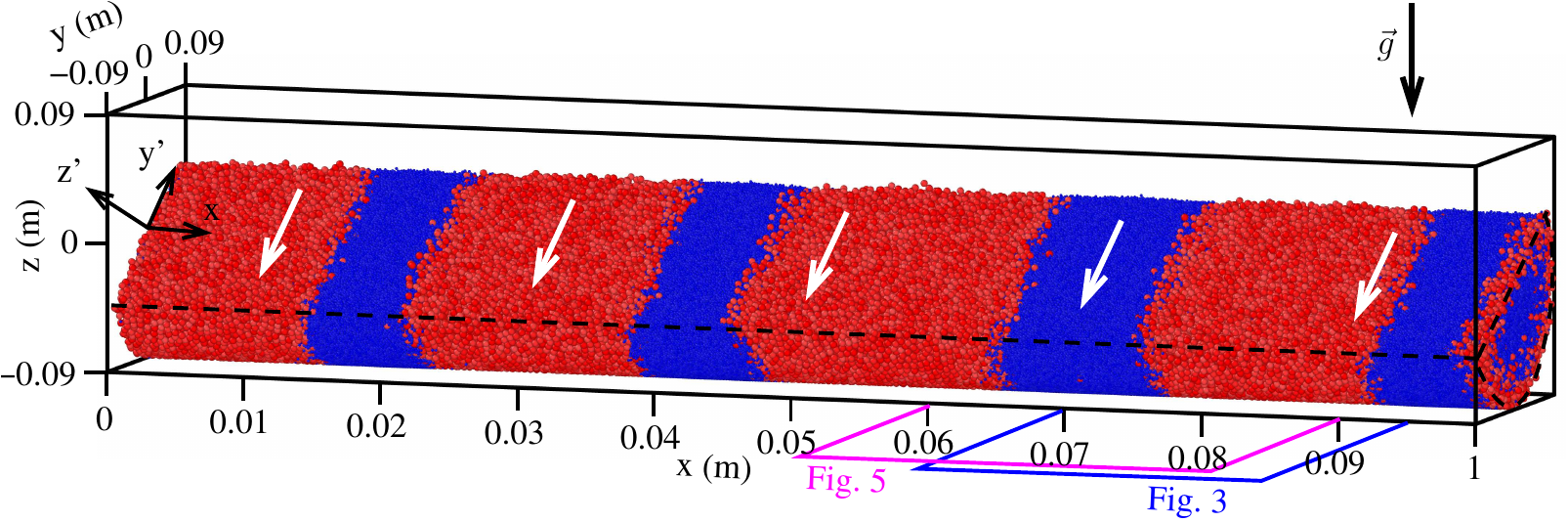}
\caption{DEM simulation of axial segregation bands in a 100~cm long, 18~cm diameter tumbler rotating at 15~rpm that is half filled
with equal volume fractions of 2~mm (blue) and 5~mm (red) spherical particles with periodic boundary conditions at 1400~s (350 rotations),
when the bands are stationary.}
\label{d3w500}
\end{figure}

Axial segregation occurs in long horizontal cylindrical rotating tumblers that are partially filled with a mixture of small and large particles. Upon rotation, particles tumble down the inclined surface in a thin flowing layer, % that is $O(10)$ particle diameters thick, 
while particles below this flowing layer are in a static zone that is in solid body rotation with the tumbler. \textcolor{black}{It is well-known that within a few tumbler rotations, radial size segregation occurs in which small particles percolate to the bottom of the surface flowing layer forcing large particle to rise to the top of the flowing layer ~\cite{DonaldRoseman62, Ristow94, Clement95}.} Thus, there is a layer of large particles over a layer of small particles flowing down the inclined free surface of the granular bed, while the particles below the flowing surface layer are in solid body rotation. After many more tumbler rotations \textcolor{black}{following radial segregation,} axially spaced bands rich in small or large particles appear (Fig.~\ref{d3w500})\textcolor{black}{~\cite{Bridgwater1976, HillKakalios94, ChooMorris98, AlexanderMuzzio04}. Although first reported in 1940~\cite{Oyama40}, axial segregation did not attract attention until much later \cite{DonaldRoseman62, DasguptaKhakhar91, Nakagawa94, ZikLevine94, HillKakalios94, HillKakalios95, ChooMorris98, Levine99, Rapaport02, AlexanderMuzzio04}, and it remains unexplained. The axial} bands typically have a wavelength of approximately one tumbler diameter~\cite{CharlesMorris05, FiedorOttino06, JuarezLueptow08, JuarezOttino10}, although they eventually tend to merge and coarsen \cite{Frette97, HillCaprihan97, ChooMorris97, ChooMorris98, FiedorOttino03, FiedorOttino06, JuarezLueptow08, JuarezOttino10}. The bands can occur, or not occur, under a wide range of conditions including various fill levels, rotation speeds, particle shape and size, species concentrations, tumbler cross-sections, with submerged as well as dry particles, and in tilted tumblers \cite{JainKhakhar01, FiedorOttino03, Arndt05, JuarezLueptow08}. When observed using an index-matched interstitial fluid or magnetic resonance imaging, small particle bands correspond to regions where the radially-segregated core of small particles extends to the flowing layer surface; large particle bands comprise an annulus of large particles surrounding the core of small particles~\cite{HillKakalios95, JainKhakhar01, KhanMorris2004, Arndt05,TaberletLosert04,TaberletNewey06, ChenLueptow11}, evident at the right end of the domain in Fig.~\ref{d3w500}.

\FloatBarrier

Proposed mechanisms for axial bands include differences in the angle of repose~\cite{ZikLevine94, HillKakalios94, HillKakalios95} or mobility of particle species~\cite{ZikLevine94, Levine99}, and the concept of ‘negative diffusivity’ has been used to rationalize axial segregation~\cite{HillKakalios94, HillKakalios95, Aranson1999a, Aranson1999b}. However, these concepts address how bands might be reinforced, not their origin~\cite{AlexanderMuzzio04, KhanMorris2004, ChenLueptow11} nor the subsurface bulges in the radially segregated core of small particles that precede visible bands at the surface~\cite{HillCaprihan97, ElperinVikhansky1999, AlexanderMuzzio04}. What is known is that  bands of large particles typically first appear near the tumbler endwalls \cite{DonaldRoseman62, Bridgwater1976, HillKakalios94, FiedorOttino03, Arndt05, Rapaport07, ChenLueptow10, ChenLueptow11} due to the interaction between endwall friction and radial segregation~\cite{DonaldRoseman62, PohlmanOttino06, SantomasoOlivi04, ChenOttino08, ChenLueptow11, DOrtonaThomas18}. However, the initial band formation at the endwalls does not explain bands that initiate near the middle of a tumbler~\cite{ChenLueptow11} or bands that form with periodic endwall conditions (Fig.~\ref{d3w500})~\cite{ChenLueptow11}.

Here, we propose a mechanism for axial segregation and band formation that is related to the recent discovery of a granular flow instability~\cite{DOrtonaThomas20} that is analogous to the fluid RT instability occurring when a static dense fluid is atop a static less dense fluid. A hint that the granular RT instability may be connected to axial band formation in rotating tumblers is that bands form more readily for mixtures when the large particles are slightly more dense than small particles~\cite{AlexanderMuzzio04, Kondo2023}. Indeed, we demonstrate here that a granular RT instability induces band formation of size-bidisperse iso-density particles in rotating tumblers, 
and the segregated axial bands of particles are maintained by the emergence of streamwise counter-rotating rolls driven by size segregation.
{ The excess density necessary for a RT instability is a consequence of a property of granular materials: a polydiserse granular mixture can pack more densely than monodisperse particles and, hence, has a greater bulk density~\cite{LochmannOger06,WeaireAste08}. A RT instability
resulting from the bulk higher density of mixed polydisperse particles has, to our knowledge, never been reported for granular flows.}
%\section{Numerical model}

\emph{Simulations}---The Discrete Element Method (DEM) code  LIGGGHTS~\cite{CundallStrack79,Kloss2012} is used to explore size-bidisperse granular flow in a smooth frictional tumbler wall having inner diameter $D=2R=18$~cm and length $L=100$~cm, using periodic boundary conditions at the ends of the computational domain \textcolor{black}{ to avoid any frictional effects that would occur if frictional endwalls were used}.
Tumblers are half-filled with equal volumes of small (diameter $d=2$~mm) and large particles (various diameters $d_l>2$~mm). % such that the average particle volume fraction in the tumbler is $\bar{f}_s=\bar{f}_l=0.5$, where subscripts $s$ and $l$ refer to small and large particles, respectively. % The two particle species have sizes $d=2$~mm and $d_l$, where $d_l>2$~mm is adjusted to vary the particle size ratio. %In some cases, the particles have different densities such that the large particle density is $\rho_l = R_\rho \rho$, where $\rho$ is the density of the small particles. 
Particle properties correspond to cellulose
acetate: density $\rho =$ 1308~kg~m$^{-3}$, restitution coefficient $e = 0.87$ \cite{DrakeShreve86,FoersterLouge94,SchaferDippel96}, friction coefficient $\mu = 0.7$, and uniform size distributions of 0.95-1.05 times $d$ or $d_l$. %The total number of particles 
%ranges from about $9 \times 10^4$ to $1 \times 10^6$. % \textcolor{blue}{(is this the right number of particles?)}.
%To avoid a close-packed structure, the particles have a
%uniform size distribution ranging from 0.95$d$ to 1.05$d$. % \textcolor{blue}{(is the distributions uniform and is the size range correct?)}. 
%The friction
%coefficient between particles and between particles and walls is set to 
%$\mu = 0.7$. 
%Gravitational acceleration is $g$ = 9.81~m~s$^{-2}$. 
%, the collision time is $\Delta t$ =10$^{-4}$ s, consistent with previous simulations \cite{TaberletNewey06,ChenLueptow11,ZamanDOrtona13} and sufficient for modeling hard spheres \cite{Ristow00,Campbell02,SilbertGrest07}.  These parameters correspond to a stiffness coefficient $k_n = 7.32\times 10^4$ (N m$^{-1}$) \cite{SchaferDippel96} and a damping coefficient $\gamma_n = 0.206 $~kg~s$^{-1}$. 
Particle interactions are calculated using a linear contact model with Young's modulus $E=10^7$ Pa and Poisson's ratio $\nu=0.3$. The
integration time step is $5\times 10^{-6}$~s to provide stable results.
Particles are initially randomly
distributed within the tumbler volume at zero gravity. Then gravity ($g$ = 9.81~m~s$^{-2}$) and tumbler rotation are turned on simultaneously so that the mixed particles fall to the bottom of the tumbler and flow due to tumbler rotation at %$\omega=3$ to 30 rpm (Froude number between $0.0009\le Fr=\omega^2 R/g\le 0.09$) 
$\omega=15$~rpm (Froude number $Fr=\omega^2 R/g=0.023$)
in a steady flowing layer at the free surface that is essentially flat above a static region that is in solid body rotation.  

%The velocity throughout the entire domain is obtained by binning particles in a 3D grid and averaging their velocity over 50~s of physical time (2.5 $\times$ 10$^7$ integration time steps) to assure an adequately smooth velocity field.

%\section{Long tumblers}

%Our initial objective is to find the conditions for which the system is most unstable, that is, the conditions for which the small and large particle bands form most quickly and are maximally segregated. 
\emph{Segregation time evolution}---To quantify the evolution of the degree of axial segregation, we compute $I_\mathrm{seg}(t)=(1/L)\int_0^L |\bar{f_l}-f_l(x,t)|+|\bar{f_s}-f_s(x,t)|{\rm d}x$, where $f_i(x,t)$ is the volume fraction for species $i$ averaged across the $y-z$~plane in the particle bed at axial position $x$ and time $t$, and the average species volume fraction in the tumbler is $\bar{f}_s=\bar{f}_l=0.5$, where subscripts $s$ and $l$ refer to small and large particles, respectively.  $I_\mathrm{seg}=0$ corresponds to no axial segregation when the two particle species are completely mixed or when there is an axially-invariant radial core of small particles.
%surrounded by a periphery of large particles.
Perfect axial segregation with axial bands of pure small or pure large particles corresponds to \mbox{$I_\mathrm{seg}=1$}. Note that for most axial segregation experiments it is only the appearance of bands at the surface or through clear tumbler walls that is observed~\cite{Frette97, ChooMorris97, FiedorOttino03, JuarezLueptow08}, whereas $I_\mathrm{seg}$ measures segregation that can manifest as subsurface bulges in the radial segregation core~\cite{HillKakalios95, JainKhakhar01, Rapaport02, KhanMorris2004, Arndt05,TaberletLosert04,TaberletNewey06, ChenLueptow11}. The time evolution of $I_\mathrm{seg}$ indicates how quickly axial segregation occurs, and the value at which $I_\mathrm{seg}$ plateaus indicates the steady-state degree of axial segregation. %Note that in most experiments on axial segregation it is only the appearance of bands at the surface or clear tumbler walls that is observed~\cite{Frette97, ChooMorris97, FiedorOttino03, JuarezLueptow08}, whereas $I_\mathrm{seg}$ measures segregation that can manifest as subsurface bulges in the radial segregation core~\cite{HillKakalios95, JainKhakhar01, Rapaport02, KhanMorris2004, Arndt05, TaberletNewey06, ChenLueptow11}.

\begin{figure}[hptb] %fig 2 equivalent to 498 words
\includegraphics[width=\linewidth]{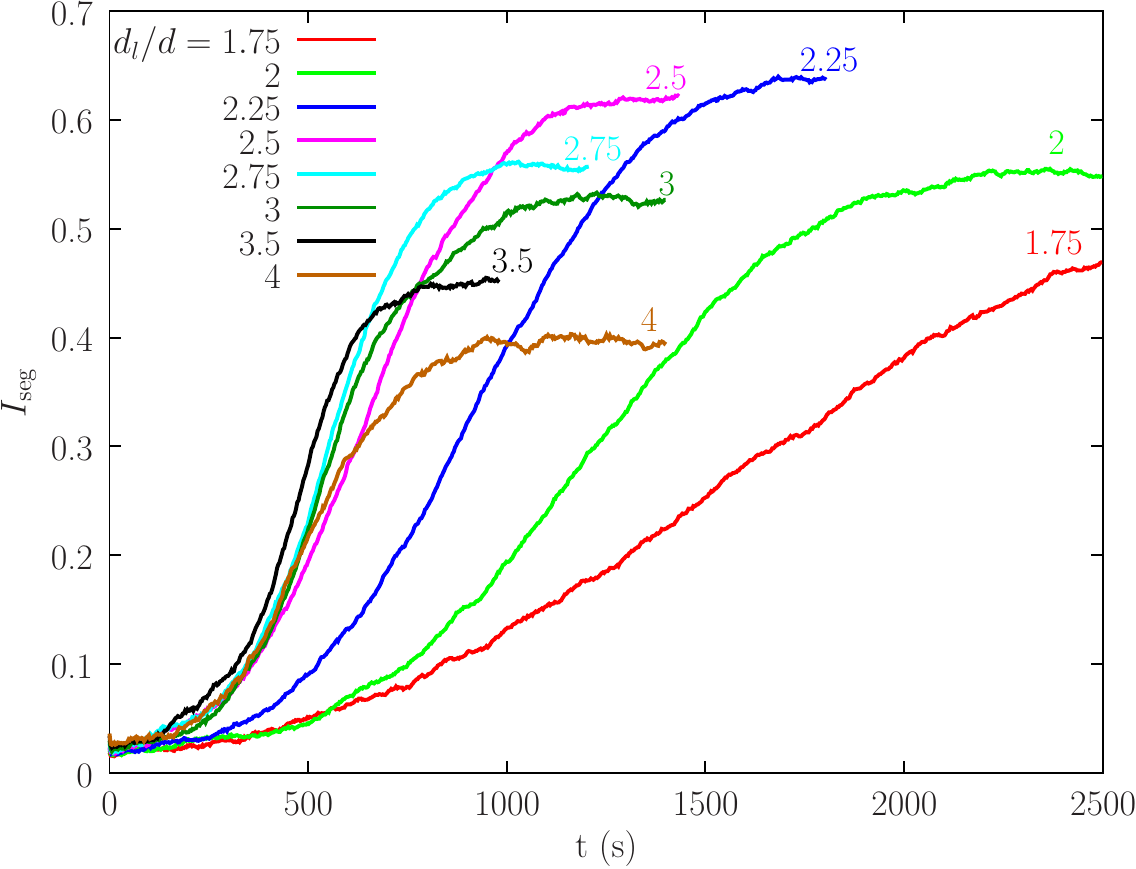}% equiv 306 words, final figure
\caption{%Time evolution of $I_\mathrm{seg}$ for particle size ratios $d_l/d=1.75$ to 4 with $d=2$~mm. The tumbler is 1~m long ($L/d=500$) with periodic end conditions, 18~cm in diameter ($D/d=90$), filled to 50\% with equal volume fractions of small and large particles, and rotating at 15~rpm. 
% SHORT VERSION %labels have been added
Evolution of $I_\mathrm{seg}$ for $1 < d_l/d \leq 4$ and $L/d=500$.
}
\label{isvsdl}
\end{figure}

Although we consider a wide range of conditions in a companion paper,
%that is under preparation, 
here we provide an example of how the particle size ratio, $d_l/d$, affects the evolution of the axial segregation, $I_\mathrm{seg}$, for $D/d=90$  (Fig.~\ref{isvsdl}). In all cases, $I_{\rm seg}$ evolves approximately exponentially at first, followed by linear growth before it %$I_{seg}$ %starts to flatten as it reaches its plateau. 
%$I_{seg}$ grows quickly before it 
flattens and reaches a plateau. %The time evolution of axial segregation is much faster for size ratios $d_l/d\geq2.5$ than for smaller size ratios, consistent with previous experiments~\cite{AlexanderMuzzio04}. 
The growth rate increases with $d_l/d$, consistent with previous experiments~\cite{AlexanderMuzzio04}, and saturates for $d_l/d \geq 2.5$.
%Of these "fast" cases, the $d_l/d=2.5$ case reaches a high plateau most quickly, and, hence, we consider this case in more detail shortly. 
The plateau for $I_\mathrm{seg}$ is highest for $d_l/d=2.25$ and 2.5 (the case we consider in detail), a consequence of the lower propensity for size segregation at smaller and larger $d_l/d$~\cite{Schlick15}. %the time evolution is slower and
%For $d_l/d>2.5$, the segregation evolution is fast, but the segregation is not as strong, with a lower plateau for $I_\mathrm{seg}$. %In all cases, the initial segregation evolution is approximately exponential at first, followed by linear growth before $I_{seg}$ starts to flatten as it reaches its plateau.
%Of these cases, the $d_l/d=2.5$ case reaches a high plateau most quickly, and, hence, we consider this case in detail.

\emph{Band formation mechanisms}---We propose that after initial radial segregation of large and small particles, axial segregation comes about due to a granular RT instability in the flowing layer much like that for a two-layer flow of large dense particles above small light particles~\cite{DOrtonaThomas20}. The axially segregated pattern is maintained by streamwise recirculation rolls that repeatedly redistribute the large and small particles % into concentrated bands of each species while passing through the flowing layer.
mixed by diffusion at the interface into their own band.

\begin{figure}[hptb] %fig 3 eq to 950 words
\includegraphics[width=\linewidth]{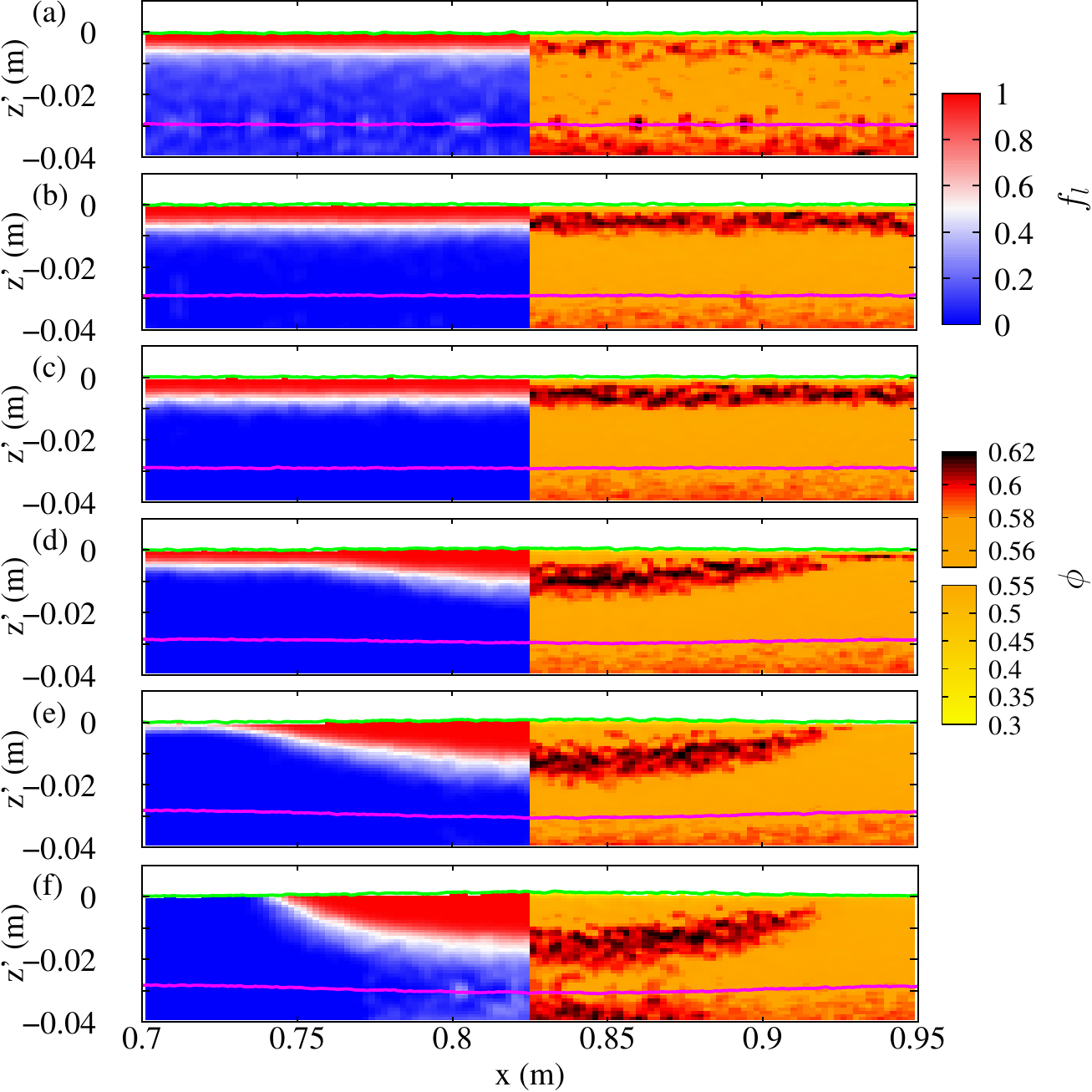}%equiv 566 words
%caption of figure3b
%\caption{Large particle species fraction, $f_\mathrm{l}$, and particle volume fraction, $\phi$, at $t=50$~s (a, b), at $t=440$~s (c) and at $t=1300$~s (d, e) in the $x-z'$~plane at $y'=0$ for a half-filled 100~cm long, 18~cm diameter cylindrical periodic tumbler rotating at 15~rpm with equal volumes of 2-mm (blue) and 5-mm (red) spherical particles ($d_l/d=2.5$). Only the flowing region ($0\lesssim z'\lesssim 0.04$~m) and a portion of the tumbler ($0.7<x<0.95$~m) that includes a complete large particle band and two half small particle bands is shown. Green lines indicate the free surface ($\phi=0.3$), and magenta lines indicate the bottom of the flowing layer (zero velocity in the laboratory reference frame).
%caption of figure3bb
%LONG OLD CAPTION
%\caption{Large particle species fraction, $f_\mathrm{l}$, and particle volume fraction, $\phi$, at $t=50$~s (a, b), at $t=440$~s (c, d) and at $t=1300$~s (e, f) in the $x-z'$~plane at $y'=0$ for a half-filled 100~cm long, 18~cm diameter cylindrical periodic tumbler rotating at 15~rpm with equal volumes of 2-mm (blue) and 5-mm (red) spherical particles ($d_l/d=2.5$). Only the flowing region ($0\lesssim z'\lesssim 0.04$~m) and a portion of the tumbler ($0.7<x<0.95$~m) that includes a complete large particle band and two half small particle bands is shown. Green lines indicate the free surface ($\phi=0.3$), and magenta lines indicate the bottom of the flowing layer (zero velocity in the laboratory reference frame).
%SHORT CAPTION
\caption{Large particle species fraction, $f_\mathrm{l}$ (red-blue colorbar), and particle volume fraction, $\phi$ (yellow-black colorbar), at \textcolor{black}{(a) $t=15$~s, (b) $t=30$~s,} (c) $t=50$~s, (d) $t=500$~s, (e) $t=700$~s and (f) $t=1250$~s in the $x-z'$~plane at $y'=0$ ($d_l/d=2.5$). Only the flowing region ($0\lesssim z'\lesssim 0.04$~m) is shown. Green lines indicate the free surface ($\phi=0.3$), and magenta lines indicate the bottom of the flowing layer.}
%\textcolor{blue}{1. Would it be possible to indicate the approximate location of the bottom of the flowing layer based on zero velocity? 2. Would it be possible to add the axes like in Fig.~\ref{dispmap}?}
%{\red (e) and (f) same as fig (a) and (b) but with complete static zones, we have to choose. 
%(a and b will improve, I had no more the data, so I'm regenerating them.
%I will add (a), (b)... }
\label{color}
\end{figure}

The distributions of large and small particles ($d_l/d=2.5$)  in the flowing layer and the overall particle volume fraction, $\phi$, are shown in Fig.~\ref{color} for %different times in the formation of the bands. 
\textcolor{black}{radial segregation and subsequent formation of the bands. Radial segregation occurs quickly after the start of rotation with large particles (red) at the flowing layer surface and small particles (blue) segregated below them within $t=15$~s (3.75 tumbler rotations, Fig.~\ref{color}(a)), consistent with previous results~\cite{DonaldRoseman62, Ristow94, Clement95,Bridgwater1976, HillKakalios94, ChooMorris98, AlexanderMuzzio04}. As radial segregation progresses, {which corresponds to segregation normal to the free surface in the $z'$-direction}, the large and small particle layers become more concentrated ($t=30$~s, Fig.~\ref{color}(b)), so that by $t=50$~s (12.5 tumbler rotations) { the { $z'$-direction} segregation reaches a quasi-stationary state,} {in which the segregation between the two species is incomplete}
%the particles have become fully radially segregated  
(Fig.~\ref{color}(c)). The mixed particle interface (the white color corresponds to a 50:50 mixture) between the large and small particle layers results from a balance between size segregation and 
%collisional diffusion, 
{mixing.  The  balance between segregation and diffusive mixing allows the analytical solution of
the vertical particle fraction profile for a steady state segregated bidisperse chute
flow~\cite{Larcher13, SchlickInter15, Gray18} or simple shear flow~\cite{Fry19}, and, thus, the thickness of the mixing zone.
%The vertical profile of $f_l$ and the extent of the mixed layer have been predicted analytically using an advection-diffusion-segregation model for steady-state bidisperse chute flow~\cite{SchlickInter15}, 
However, a similar analysis is difficult for} tumbler flow because particles continually enter and exit the flowing layer as the tumbler rotates.} 
%At $t=50$~s (12.5 tumbler rotations) after the start of rotation, the particles have become radially segregated  (Fig.~\ref{color}(a)) with large particles (red) at the flowing layer surface and small particles (blue) below them. The white region corresponds to a 50:50 mixture resulting from a balance between size segregation and collisional diffusion. 
%The maximum of volume fraction $\phi=0.62$ corresponds to 

{The right part of Fig.~\ref{color} presents the particle volume fraction $\phi$. As radial segregation progresses, a dense layer forms around $z'\simeq-0.01$~m corresponding to the mixed particle region. At the quasi-stationary state (Fig.~\ref{color}(c)),}
the particle volume fraction  %, in Fig.~\ref{color}(b) 
varies with depth from $\phi\approx0.55$ (orange) for pure large particles at the surface, to $\phi\approx0.62$ (black) for mixed particles (which pack more densely than monodisperse particles), and then to $\phi\approx0.55$ (orange) for pure small particles. %The layer of higher $\phi$ (black) comes about because mixtures %, the layer corresponding to $0.4\lesssim f_l \lesssim 0.6$ (light blue/white/pink) in Fig.~\ref{color}(a), pack more tightly than monodisperse particles. 
The maximum of $\phi\approx 0.62$ corresponds to $f_l\approx 0.72$ (pink on the colorscale), evident as the small vertical shift between the black and white regions in Fig.~\ref{color}(c). %\sout{{\blue This maximum of density is obtained for particle fractions equivalent to those obtained in static packing \cite{LochmannOger06}.}}

We contend that for particles of the same density, the layer of mixed particles with $\phi\approx0.62$ above the layer of small particles with $\phi\approx0.55$ can provide a sufficient difference in the bulk densities %between the two layers 
for the granular RT instability to occur. Figure~\ref{color}(c-f) shows the evolution of the RT instability beginning with a downward protrusion of mixed particles at $t=500$~s (d) and an upward protrusion of small particles that has broken the free surface by $t=700$~s (e), thereby forming stationary axially segregated bands at $t=1250$~s (f). %We have confirmed that 
\textcolor{black}{Both the degree of segregation and the amplitude of a sinusoidal fit to the waviness of the mixed particle layer (white on the left side of Fig.~\ref{color}) have been measured to grow} exponentially until the small particles break the surface, as expected for a RT instability~\cite{Chandrasekhar61,Charru}. This process is unusual compared to most RT instabilities in that it occurs for a high density layer linked to the mixture that is created by diffusion and loses its excess density by segregation.
%The segregated steady-state bands that form by $t=1300$~s are shown in Fig.~\ref{color}(e). 
%\sout{The small particle band (blue) extends from the free surface to the small particle core, while the large particle band (red) is annular, surrounding a core of mainly small particles (light blue), consistent with previous results } %showing a core of small particles that periodically protrudes to the flowing layer surface~
%\sout{Plumes are asymetrical, leading to plain band of small particles, and annulus of large ones}
The resulting small particle band (blue) extends from the core to the free surface with the annular large particle band (red) surrounding the core, consistent with previous results
~\cite{HillKakalios95, JainKhakhar01, KhanMorris2004, Arndt05, TaberletNewey06, ChenLueptow11}. %High $\phi$ associated with mixed particles is evident at the interface between the small particle band and large particle bands in Fig.~\ref{color}(f).

%figure 4
\begin{figure}[hptb]
\includegraphics[width=\linewidth]{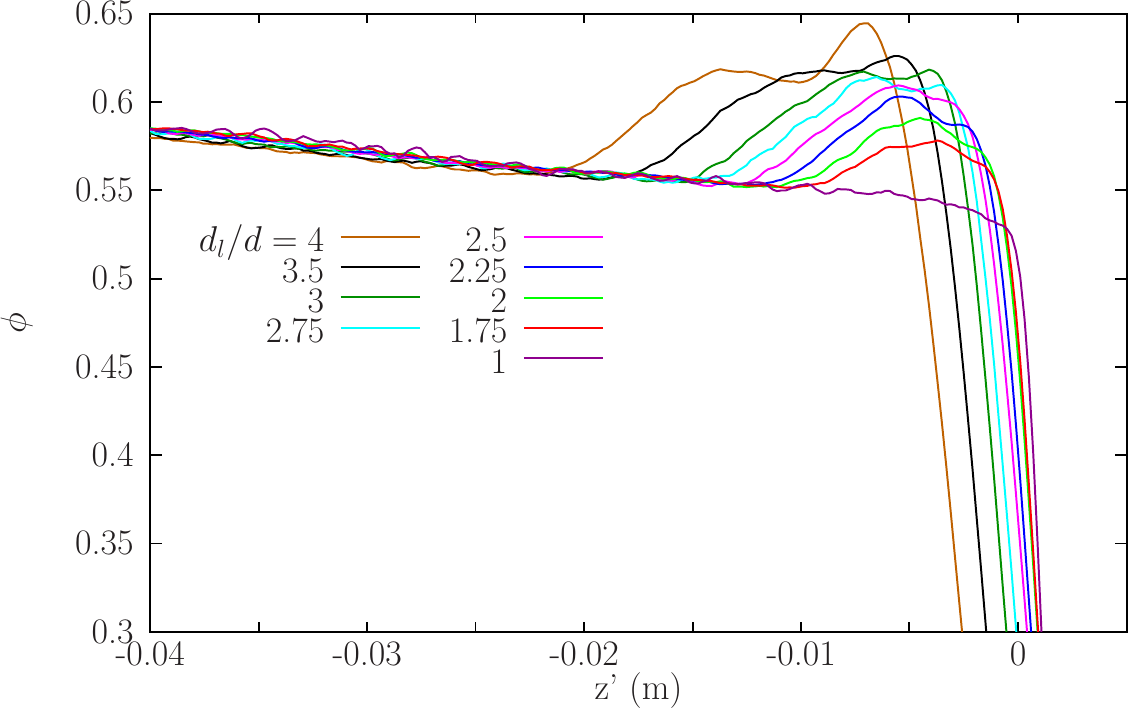}
%\includegraphics[width=\linewidth]{compmatch500+rot.pdf}%equiv 502 words
%\caption{Depthwise profile of the particle volume fraction $\phi$ halfway down the flowing layer when axial band formation is suppressed by using a {\green short} periodic 10~cm long tumbler ($L/d=50$) that is 18~cm in diameter, rotating at 15~rpm, and half-filled with equal volume fractions of small ($d=2$~mm) and large particles ($d_l/d=1$ to 4). Inset: data for $d_l/d=2$ and $\omega=3$ to 30~rpm.} %\textcolor{blue}{The slice is $-0.005\le y' \le 0.005$~cm, but how thick is it in the x-direction (axial direction)?} %4. Have I correctly described this as the $x-z'$ plane at $y'=0$? 
%{\green udo: the whole cylinder. these are narrow cylinders to prevent the formation of axial segregation. so I can use the whole length $W=50d$. I changed the figure to include as an inset the effect of the rotation speed.} 
 %(b) Volume fraction of small particles (dotted curves), of large particles (dashed curves) and the sum (solid curves) for three size ratios ($d_l/d=1.5$, 2.5 and 3.5).
% {\red Nathalie: there is no more (b) and the inset does not correspond.}
%SHORT CAPTION
\caption{Depthwise profile of the particle volume fraction $\phi$ at $y'=0$ for $1 \leq d_l/d \leq 4$ when axial band formation is suppressed by using a short 10~cm long tumbler ($L/d=50$).}
\label{volfrac}
\end{figure}
%\FloatBarrier

%\sout{The differing densities of the layers can also be shown by considering the depthwise profile of $\phi$ after radial segregation is established but before bands form (Fig.~\ref{volfrac}). Band formation is suppressed by using a short periodic tumbler with $L/d=50$, which is well below the length required for band formation~\cite{CharlesMorris05, FiedorOttino06, JuarezLueptow08, JuarezOttino10}. }%(REF NEEDED).

%The depthwise profile of $\phi$ after radial segregation is established but before bands form is shown in Fig.~\ref{volfrac} by using a tumbler too short for bands to form~\cite{CharlesMorris05, FiedorOttino06, JuarezLueptow08, JuarezOttino10}. 
Depthwise profiles of $\phi$ after radial segregation is established but before bands form demonstrate how $\phi$ for the mixed layer depends on $d_l/d$. Figure~\ref{volfrac} presents smoothed profiles of $\phi$ obtained using a tumbler too short for bands to form~\cite{CharlesMorris05, FiedorOttino06, JuarezLueptow08, JuarezOttino10} allowing a long integration time ($50\le t \le 100$~s).
%{\red Smoothed depthwise profiles of $\phi$ after radial segregation is established but before bands form, obtained using a tumbler too short for bands to form~\cite{CharlesMorris05, FiedorOttino06, JuarezLueptow08, JuarezOttino10} \sout {allowing a long integration time ($50\le t \le 100$~s)}, demonstrate how $\phi$ for the mixed layer depends on $d_l/d$ (Fig.~\ref{volfrac}).}
% (12.5 to 25 rotations). 
%Figure~\ref{volfrac} shows the depthwise ($z'$) 
Profiles of $\phi$ are taken halfway down the flowing layer averaged over a 1~cm thick slice in the $x-z'$ plane. %  and over $50\le t \le 100$~s %(12.5 to 25 rotations) 
%for several values of $d_l/d$.   %This corresponds to the average $z'-$variation in $\phi$ at the plane shown in Fig.~\ref{color}(c).  
The bottom of the flowing layer is at $z'\approx -0.04$~m, and the free surface is at $z'\approx 0$, where $\phi$ decreases sharply. For all size ratios, there is a region around $z'=-0.005$~m where $\phi$ is significantly larger, indicating a denser layer of mixed particles above a less dense layer of monodisperse small particles. 
%For example, for $d_l/d=2.5$, $\phi\approx 0.61$ at $z'=-0.005$, while $\phi \approx 0.55$ at $z'=-0.013$, deeper in the flowing layer. The ''bump" in the curves in Fig.~\ref{volfrac} directly corresponds to regions where the particles are mixed, rather than monodisperse. 
The excess in bulk density, %of these two layers resulting from the difference in $\phi$, 
typically around 10\%, is what drives the RT instability, even without any density difference between the particle species. %$\phi$ decreases sharply at $z'\approx 0$, corresponding to %that have segregated to the surface of the flowing layer the free surface.

%It is evident in Fig.~\ref{volfrac} that 
Figure~\ref{volfrac} shows that $\phi$ in the mixed layer %of particles
increases with increasing $d_l/d$, as would be expected for more efficient packing at higher $d_l/d$. %This drives the faster band formation with increasing $d_l/d$ %, evident in (Fig.~\ref{isvsdl}). %{\green To analyse in more details, the maximum of the volume fraction 
For $d_l/d=1$ to 2.5, the maximal $\phi$ near the surface increases with a bump that is relatively narrow, but for $d_l/d=2.5$ to 4, the maximum only increases slightly while the bump broadens. Referring back to Fig.~\ref{isvsdl}, these two ranges correspond to the exponential growth strongly increasing from $d_l/d=1.75$ to 2.5 but saturating for larger $d_l/d$. The increase of the growth rate with an increase in the density difference is expected for a RT instability~\cite{Chandrasekhar61,Charru}.  

%The inset in Fig.~\ref{volfrac} showing the $\phi$ profile for {\green $d_l/d=2$} at different tumbler rotation speeds amplifies that a gradient in $\phi$ is a necessary condition for bands to form. When the gradient of $\phi$ in the flowing layer is too small, such as for 30 rpm, bands do not form. \sout{On the other hand, when the gradient in $\phi$ is too large, as is the case for 3 rpm, bands also do not form, possibly because the higher density layer is too thin to support the instability.} {\green But for small rotation speeds ($\omega=3$ and 5~rpm), bands do not also form, although the density gradient is strong. In these cases, large particles organise in 2 layers at the free surface, characteristic of a strong size segregation that is expected at low velocity. Segregation maintains a flat dense upper layer preventing the RT instability.}

%The segregated steady-state bands that form by $t=1300$~s are shown in Fig.~\ref{color}(c). The small particle band (blue) extends from the free surface to the small particle core, while the large particle band is annular, with a core of small (blue) or mixed (white) particles below it, consistent with previous experimental results showing a core of small particles that periodically protrudes to the flowing layer surface~\cite{HillKakalios95, JainKhakhar01, KhanMorris2004, Arndt05, TaberletNewey06, ChenLueptow11}. High $\phi$ associated with mixed particles is evident at the interface between the small particle band and large particle bands in Fig.~\ref{color}(d).

%figure 5 equivalent to 640words
\begin{figure}[hptb]
\includegraphics[width=\linewidth]{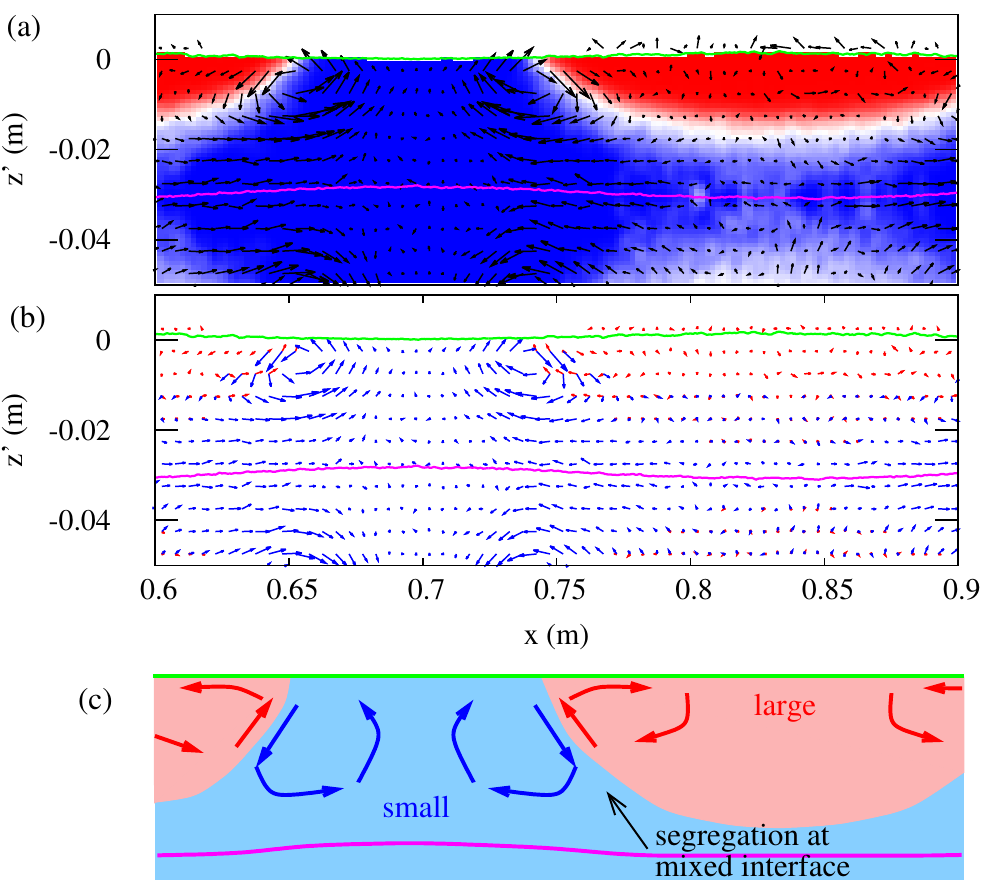}%equiv 534words, it is the final figure
\caption{Displacement vector maps (magnified 15 times) for (a) both species, and (b) each species multiplied by its volume fraction, $f_\mathrm{i}\phi$, in the $x-z'$ plane %halfway down the flowing layer
($y'=0$) for fully developed axial segregation at $t=1300$~s. %in a periodic 100~cm long 18-cm diameter cylinder filled to 50\% with with equal volumes of 2-mm (blue) and 5-mm (red) spherical particles ($d_l/d=2.5$). Vectors have been magnified 15 times; 
Colors in (a) correspond to $f_l$ according to Fig.~\ref{color}; red and blue vectors in (b) correspond to large and small particles, respectively. %Thick red and blue arrows indicate the sense of rotation. 
(c) Sketch of recirculation rolls in the large (red) and small (blue) particle regions. Green lines corresponds to the free surface and magenta lines to the lower limit of the free flowing layer.%; (b) the large (red) and small (blue) particles; %{\red (c) and (d) same as (a)  and (b) for only the flowing zone.}
}
\label{dispmap}
\end{figure}
\emph{Recirculation rolls} ---
%version Feb 8
The usual ascending and descending plumes expected for both the granular and fluid RT instabilities that lead to an inversion of the two phases do not occur here. The descending plumes do not spread at the lower bound of the flowing layer but remain as fixed downward protrusions (Fig.~\ref{color}(f)). The ascending plumes are not symmetric with the descending plumes. %\sout{They are composed of nearly pure small particles and, as a consequence, are less dense than the mixture of large and small particles at the head of the descending plumes. }
Ascending plumes may reach the free surface and form bands of pure small particles as observed here and in experiments or stay below the free surface in the case of subsurface bulges. Neither the ascending  nor the descending plumes spread in the spanwise direction at the surface or at the bottom of the flowing layer as would occur in the fluid RT instability.
%\sout{\green We observed that the downwise plumes %of large particles 
%do not reach the bottom of the flowing layer, but remains in a frozen state (see Figs.~\ref{color}(e-f)). This is unusual in a RT instability where the two fluids usually reverse completely. %The mechanism that stops the sinking plumes is analogous than at low rotation speed. The head of the plume reach a region where the flow velocity is lower, the pressure is higher, the size segregation becomes more efficient and block the RTI. The ascending plumes are not symetrical and reach the free surface, but can not spread any further due to the frozen descending plumes.} 
%{\blue The usual upward and downward plumes expected for both the granular and fluid RT instability that eventually reverse direction at the top and bottom of the domain do not occur here.} 
Instead, the combination of RT instability and segregation creates
recirculation rolls in the $x-z'$ plane %where the flow is normal to the plane of the image
(Fig.~\ref{dispmap}), shown as particle displacement vectors over a complete circuit through the flowing layer and solid body rotation based on the averaged velocity field. % make the longitudinal rolls related to the instability evident. %Fig.~\ref{dispmap} displays the average particle displacement with each circuit through the flowing layer for steady-state bands at $t=1000$~s (250 tumbler rotations). To obtain the displacement vectors, the velocity throughout the entire domain at steady-state is obtained by binning particles in a 3D grid and averaging their velocity over X s \textcolor{blue}{What is the time used here?} to assure an adequately smooth velocity field. Mean particle trajectories are obtained by integrating the particle velocity based on the velocity field.
%or by averaging particles trajectories based on particle positions stored every 0.1s. \textcolor{blue}{Which of these two approaches is used here?} udo: I used the velocit field, particle trajectories are not accurate enough
%The vector map is based on the displacement in the particle mean trajectories starting from the $y'=0$~plane, making a complete circuit through the solid body rotation zone, and returning to the $y'=0$~plane again.
 Figure~\ref{dispmap}(a) shows the vector displacement map for all particles %, large and small, for the same conditions as in Fig.~\ref{color}(e), but with the domain slightly shifted axially to focus on the small particle band. T
overlaying the large particle volume fraction, $f_l$. % overlays the vectors to show the large and small particle bands. 
There is a pair of relatively strong streamwise-oriented recirculation rolls on either side %of the small particle band centered at $x\approx 0.7$~m. 
of the interface between the bands.
%The recirculation rolls in the flowing layer are somewhat different from those for the granular RT instability with a layer of dense large particles above a layer of light small particles in a chute flow~\cite{DOrtonaThomas20}.  The RT instability appears to initiate the formation of the bands and likely determines their wavelength, but it does not induce Rayleigh-B\'enard-like convection rolls like those for the granular RT instability for bi-density layers in chute flow~\cite{DOrtonaThomas20}, where the rolls extend through the entire flowing layer thickness with descending large-dense particle plumes and ascending plumes of mainly small-light particles. Instead, the rolls associated with axial bands in a tumbler are in the upper part of the flowing layer, where the interface between the bands is tilted. The resulting large particle bands are also different in that they are wider and have an underlying core of small particles. 

The ongoing vortical motion is decomposed into separate displacement vector maps for large and small particles in Fig.~\ref{dispmap}(b) based on their individual velocity fields, %calculated separately for each species and 
multiplied by the species volume fraction, $f_\mathrm{i}\phi$. %, to account for low species concentration in certain regions. 
Areas with only one color of vectors have only one of the species present. The two species follow different paths: the small particle recirculation roll centered at $x\approx 0.74$~m %, it is clear that the downward motion of small particles in the recirculation roll 
occurs in the mixed and small particle regions (Fig.~\ref{dispmap}(a)). %In the same region, large particles move upward. 
The relative motion between the species is due to size segregation. As sketched in Fig.~\ref{dispmap}(c), the downward segregation of small particles at the mixed tilted region drives the small particle recirculation roll. %\sout{, which brings small particles to the surface at $0.70 \lesssim x \lesssim 0.74$~m.} 
At the same time a much weaker large particle recirculation roll (red) %\sout{in the region $0.74 \lesssim x \lesssim 0.83$~m }
rotates in the same direction, driven by the upward segregation of the large particle in the tilted mixed region. The difference in the strength of the recirculation cells is likely due to the denser mixed region inducing a downward motion that favors the small particle recirculation rolls.

%Nathalie proposes :
%Here, the RT instability appears to initiate the formation of the bands and likely determines their wavelength, but the recirculation rolls are different from the Rayleigh-B\'enard-like convection rolls that extend through the entire flowing layer thickness with descending large-dense particle plumes and ascending plumes of mainly small-light particles that occur for the RT instability with a layer of dense large particles above a layer of light small particles in a chute flow~\cite{DOrtonaThomas20}. 
%Umberto: the previous sentence is long, complex a repetitive, what about that one
Here, the RT instability appears to initiate the formation of the bands and likely determines their wavelength, but the recirculation rolls are different from the Rayleigh-B\'enard-like convection rolls for flow down an incline, where rolls extend through the entire flowing layer thickness with descending large-dense particle plumes and ascending plumes of mainly small-light particles flow~\cite{DOrtonaThomas20}.
Instead, there are twice as many rolls associated with axial bands in a tumbler, and they are in the upper part of the flowing layer, where the interface between the bands is tilted. The resulting large particle bands are also different in that they are wider and have an underlying core of mainly small particles.  %{\blue Furthermore, unlike plumes in the fluid RT instability or its granular analog, here the downward motion of the plume is interrupted by the combined effects of the lower bound of the flowing layer, diffusion, and segregation.}%, and opposes to the large particle ones. 
%Proposed version Feb 8
Furthermore, the downward motion of descending plumes is controlled by the combined effects of diffusion, which favors mixing, and the subsequent increase in $\phi$, and segregation, which reduces mixing. As a plume sinks, its head reaches a region where $\phi$ is higher and the velocity is lower, both enhancing segregation efficiency.
%{\red Here is a second effect that is nearly not visible in fig 5, but visible in fig 13 in the PRE article. may be we should let this for the PRE.} 
 The plume stops when a balance is obtained between the density excess, mixing, and segregation, thereby maintaining a frozen pattern of depthwise protrusions and a stationary band pattern that can be visible at the surface %\sout {These mechanisms combine to block the downward plume near the lower bound of the flowing layer, inducing a high segregation index or at an intermediate level for sub-surface bulges (moderate $I_{\rm seg}$). In cases where segregation is very efficient, no instability occurs $I_{\rm seg}\approx 0$.} 

\emph{Discussion}---%{\red These results explain how the granular RT instability initiates the formation of plumes leading to axial segregation that is reinforced by pairs of segregation driven recirculation rolls on both sides of tilted regions of the interface.}
These results explain how the granular RT
instability initiates the formation of plumes leading to axial segregation. Simultaneously, pairs of segregation driven recirculation rolls develop on both sides of the tilted regions of the interface. This RT instability is unusual in the sense that  that the third denser layer that appears between the large and small particle layers results from the mixture of granular species and does not have a constant volume.  As a consequence, the final depth and shape of the bands result from a complex interaction between instability, mixing, and segregation. 
%{\green These results explain how the granular RT instability initiates the formation of plumes leading to axial segregation and recirculation rolls driven by segregation at the tilted interface between large and small particle bands.}
\textcolor{black}{Axial bands form over a range rotation speeds, particle size ratios, tumbler diameters, fill levels, and particle densities, %{\blue (also see \cite{SuppMat})}, 
although the 
 formation of bands is sensitive in varying degrees to these factors,} 
all of which influence the RT instability, the segregation process, or both, and which will be explored extensively in a detailed companion paper. \textcolor{black}{Even more intriguing are the questions of why the wavelength of the bands is approximately equal to the tumbler diameter, likely due to specific aspects of tumbler flow, and why the bands eventually coalesce to coarsen the band structure.% {\blue (see \cite{SuppMat})}. %These are 
The latter is likely a consequence of the complex interaction between the RT instability and particle segregation. Nevertheless, the current} results finally resolve the long-standing question of the basic mechanisms behind the appearance of axially segregated bands for size-bidisperse iso-density species in a rotating tumbler that has long puzzled researchers.

\emph{Acknowledgments}---%Centre de Calcul Intensif d'Aix-Marseille University is acknowledged for granting access to its high performance computing resources.
This work was performed using HPC resources from GENCI–IDRIS (Grant 2022-102451).

\end{document}